\documentstyle{article}
\begin{document}
\LARGE
\begin{center}
\bf Quantum Kaluza-Klein Cosmologies (V) \vspace*{0.7in}
\normalsize \large \rm

Zhong Chao Wu

Dept. of Physics

Beijing Normal University

Beijing 100875, P.R. China

\vspace*{0.55in}
\large
\bf
Abstract
\end{center}
\vspace*{.1in}
\rm
\normalsize
\vspace*{0.1in}

In the No-boundary Universe with $d=11$ supergravity, under the
$S_n \times S_{11-n}$ Kaluza-Klein ansatz, the only seed instanton
for the universe creation is a $S_7 \times S_4$ space. It is
proven that for the Freund-Rubin, Englert and Awada-Duff-Pope
models the macroscopic universe in which we are living must be 4-
instead of 7-dimensional without appealing to the anthropic
principle.

\vspace*{0.3in}

PACS number(s): 98.80.Hw, 11.30.Pb, 04.60.+n, 04.70.Dy


\vspace*{0.5in}

\pagebreak

\rm

\normalsize

In a series of papers [1] the origin of the dimension of the
universe was  investigated for the first time in quantum
cosmology. As far as I am aware, in the No-Boundary Universe [2],
the only way to tackle the dimensionality of the universe is
through Kaluza-Klein cosmologies. In the Kaluza-Klein model with
$d=11$ supergravity, under the $S_n \times S_{11-n}$ ansatz, it
has been shown that the macroscopic universe must be 4- or
7-dimensional. The motivation of this paper is to prove that the
universe must be 4-dimensional.

In  $d=11$ simple supergravity, in addition to fermion fields, a
3-index antisymmetric tensor $A_{MNP}$ is introduced into the
theory by supersymmetry [3]. In the classical background of the
$WKB$ approximation, one sets the fermion fields to vanish. Then
the action of the bosonic fields can be written
\begin{equation}
\bar{I}= \int \sqrt{-g_{11}}\left (  \frac{1}{2} R - \frac{1}{48}
F_{MNPQ}F^{MNPQ} + \frac{\sqrt{2}}{6\cdot (4!)^2}
\eta^{M_1M_2\cdots
M_{11}}F_{M_1M_2M_3M_4}F_{M_5M_6M_7M_8}A_{M_9M_{10}M_{11}} \right
)d^{11}x,
\end{equation}
where
\begin{equation}
F_{MNPQ} \equiv 4! \partial_{[M}A_{NPQ]},
\end{equation}
\begin{equation}
\eta^{A\cdots N} = \frac{1}{\sqrt{-g_{11}}} \epsilon^{A\cdots N}
\end{equation}
and $R$ is the scalar curvature of the spacetime with metric
signature $(-, +, +, \cdots +)$. The theory is invariant under the
Abelian gauge transformation
\begin{equation}
\delta A_{MNP} =  \partial_{[M}\zeta_{NP]}.
\end{equation}
It is also noticed that the action is invariant under the combined
symmetry of time reversal with $A_{MNP} \rightarrow -A_{MNP}$.

The field equations are
\begin{equation}
R_{MN} - \frac{1}{2}Rg_{MN} = \frac{1}{48}
(8F_{MPQR}F_N^{\;\;\;PQR} -g_{MN}F_{SPQR}F^{SPQR}),
\end{equation}
and
\begin{equation}
F^{MNPQ}_{\;\;\;\;\;\;\;\;\;;M}= \left
[\frac{-\sqrt{2}}{2\cdot(4!)^2 }\right ]\cdot \eta^{M_1 \cdots
M_8NPQ}F_{M_1\cdots M_4}F_{M_5\cdots M_8}.
\end{equation}

At the $WKB$ level, it is believed that the Lorentzian evolution
of the universe originates from a compact instanton solution, i.e.
a stationary action solution of the Euclidean Einstein and other
field equations. In order to investigate the origin of the
dimension of the universe, we are trying to find the following
minisuperspace instantons: the $d=11$ spacetime takes a product
form $S_n\times S_{11-n}$ with an arbitrary metric signature and
all components of the $F$ field with mixed indices in the two
factor spaces to be zero. In the factor space $S_n \;(n =1,2,3)$
the $F$ components must be vanish due to the antisymmetry of the
indices. Then $F$ must be a harmonic in $S_{11-n}$ since the right
hand side of the field equation (6) vanishes. It is known in de
Rham cohomology that $H^4(S_4) =1$ and $H^4(S_m) =0 \;\;(m\neq
4)$. So there is no nontrivial instanton for $n = 1,2,3$. For
$n=5,6$, both $F$ components in $S_5$ and $S_6$ must be harmonics
and so vanish. By the dimensional duality, there does not exit
nontrivial instanton either for $n= 10, 9, 8$. The case $S_4
\times S_7$ is the only possibility for the existence of a
nontrivial instanton, the $F$ components must  be a harmonic in
$S_4$, but do not have to in $S_7$. The no-boundary proposal and
the ansatz are very strong, otherwise the nonzero $F$ components
could live in  open or closed $n$-dimensional factor spaces $(
4\leq n\leq 10)$ [1].

Four compact instantons are known, their Lorentzian versions are
the Freund-Rubin, Englert, Awada-Duff-Pope and
Englert-Rooman-Spindel spaces [4][5][6][7]. They are products of a
4-dimensional anti-de Sitter space and a round or squashed
7-sphere. These spaces are distinguished by their symmetries from
other infinitely many solutions with the same $F$ field. From now
on, Greek letters run from 0 to 3 for the indices in $S_4$ and
small Latin letters from 4 to 10 for the indices in $S_7$.

One can analytically continue the $S_7$ or $S_4$ space at the
equator to form a 7- or 4-dimensional de Sitter  or anti-de Sitter
space, which is identified as our macroscopic spacetime, and the
$S_4$ or $S_7$ space as the internal space. One may naively think,
since in either case the seed instanton is the same, that the
creation of a macroscopic 7- or 4-dimensional universe should be
equally likely. However, a closer investigation shows that this is
not the case, it turns out that the macroscopic universe must be
4-dimensional, regardless whether the universe is habitable.

The Freund-Rubin is of the $N=8$ supersymmetry [4]. Here the only
nonzero $F$ components are in the $S_4$ factor space of the
instanton
\begin{equation}
F_{\mu \nu \sigma \delta} = i\kappa \sqrt{g_4}\epsilon_{\mu \nu
\sigma \delta },
\end{equation}
where $g_4$ is the determinant of the $S_4$ metric, the $F$
components are set imaginary in $S^4$ such that their values
become real in the  anti-de Sitter space, which is an analytic
continuation of the $S_4$ space, as shown below. The $F$ field
plays the role of an anisotropic effective cosmological constant,
which is $\Lambda_7 = \kappa^2/3$ for $S_7$ and $\Lambda_4 = -
2\kappa^2/3$ for $S_4$, in the sense that $R_{mn} = \Lambda_7 \;
g_{mn}$  and $R_{\mu \nu} = \Lambda_4 \; g_{\mu \nu}$,
respectively. The $S_4$ space must have radius $r_4 =
(3/\Lambda_4)^{1/2}$ and metric signature $(-,-,-,-)$, while the
$S_7$ space is of radius $r_7 =(6/\Lambda_7)^{1/2}$ and metric
signature $(+,+, \cdots +)$.

Since the metric signature of the factor space $S_4$  is not
appropriate, one has to analytically continue the $S_4$ manifold
into an anti-de Sitter space with the right metric signature
$(-,+,+,+)$. The $S_4$ metric can be written
\begin{equation}
ds_4^2= -dt^2  - \frac{3}{\Lambda_4} \sin^2\left
(\sqrt{\frac{\Lambda_4}{3}}t \right )(d\chi^2 + \sin^2 \chi
(d\theta^2 + \sin^2 \theta d\phi^2)).
\end{equation}
One can obtain the 4-dimensional anti-de Sitter space by setting $
\rho = i\chi$. However, if one looks closely in the quantum
creation scenario, this continuation takes two steps. First, one
has to continue on a three surface where the metric is stationary.
One can choose $\chi = \frac{\pi}{2}$ as the surface, set $\omega
= i(\chi - \frac{\pi}{2})$ and obtain the metric with signature
$(-,-,-, +)$
\begin{equation}
ds_4^2= -dt^2  - \frac{3}{\Lambda_4} \sin^2\left
(\sqrt{\frac{\Lambda_4}{3}}t \right )(-d\omega^2 + \mbox{cosh}^2
\omega (d\theta^2 + \mbox{cos}^2 \theta d\phi^2)).
\end{equation}
Then one can analytically continue the metric through the null
surface at $t = 0$ by  redefining $ \rho = \omega +\frac{i\pi}{2}$
and get the anti-de Sitter metric
\begin{equation}
ds_4^2= -dt^2  + \frac{3}{\Lambda_4} \sin^2\left
(\sqrt{\frac{\Lambda_4}{3}}t \right )(d\rho^2 + \mbox{sinh}^2 \rho
(d\theta^2 + \sin^2 \theta d\phi^2)).
\end{equation}

In the No-Boundary Universe, the relative probability of the
creation, at the $WKB$ level, is the exponential to the negative
of the Euclidean action of the instanton $S_7 \times S_4$
\begin{equation}
P =\Psi^* \cdot \Psi \approx \exp -I ,
\end{equation}
where $\Psi$ is the wave function of the configuration at the
quantum transition. The configuration is the metric  and the
matter field at the equator. $I$ is the Euclidean action.

If we are living in the section of the 7-dimensional de Sitter
universe with the $S_4$ space of metric (8) or the Euclidean
version of (10) as the internal space, then the Euclidean action
$I$ should take the form
\begin{equation}
I=- \int \sqrt{g_{11}}\left (  \frac{1}{2} R - \frac{1}{48}
F_{MNPQ}F^{MNPQ} + \frac{\sqrt{2}i}{6\cdot (4!)^2}
\eta^{M_1M_2\cdots
M_{11}}F_{M_1M_2M_3M_4}F_{M_5M_6M_7M_8}A_{M_9M_{10}M_{11}} \right
)d^{11}x.
\end{equation}
This is obtained through analytical continuation as in the usual
4-dimensional Euclidean quantum gravity.

However, if we are living in the section of the 4-dimensional
anti-de Sitter universe, due to the metric signature, the
Euclidean action  will gain an extra negative sign in the
continuation. This is also supported by cosmological implications.
The $R$ term in the actions can be decomposed into  $R_7 - R_4$,
where $R_7$ and $R_4$ are the scalar curvatures for the two factor
spaces with the positive-definite metric signatures. The negative
sign in front of $R_4$ is required so that the perturbation modes
of the gravitational field in the $S_4$ background would take the
minimum excitation state allowed by the Heisenberg uncertainty
principle [8]. The perturbation modes are the origin for the
structure of the Lorentzian universe in both the closed and open
models. By the same argument, if we consider 7-dimensional factor
space as our macroscopic spacetime, then one has to turn the sign
around, as the analytic continuation has taken care of
automatically.

The Euclidean action $I$ of  the $AdS_4 \times S_7$ space can be
calculated
\begin{equation}
I =\frac{1}{3}\kappa^2 V_7V_4 ,
\end{equation}
where the volume $V_7$ $\;(V_4)$ of $S_7$ $\;(S_4)$ is
$\pi^4r_7^7/3$ $\;(8\pi^2r^4_4/3)$.

The field equation (6) is derived from the action (1) for the
condition that the tensor $A_{MNP}$ is given at the boundary.
Therefore, if one uses the action (1) in the evaluation of the
wave function and the probability, then the induced metric and
tensor $A$ on it must be the configuration of the wave function.
The wave function is expressed by a path integral over all
histories with the configuration as the only boundary. In deriving
Eq. (11), one adjoins the histories in the summation of the wave
function to their time reversals at the equator to form a manifold
without boundary and discontinuity. If the configuration is given,
then one obtains a constrained instanton for the stationary action
solution. If one lifts the restriction at the equator, the
stationary action solution is a regular instanton.

The induced metric and scalar field (if there is any) at the
equator will remain intact under the reversal operation. However,
for other fields, one has to be cautious. This occurs to our
$A_{MNP}$ field. For convenience, we choose the following gauge
potential
\begin{equation}
A = i\kappa \left (\frac{3}{\Lambda_4} \right )^2 \left(\sin \left
(\sqrt{\frac{\Lambda_4}{3}} \tau \right ) - \frac{1}{3}\sin^3
\left (\sqrt{\frac{\Lambda_4}{3}} \tau \right )+ \frac{2}{3}
\right ) \sin^2 \chi \sin \theta d\chi \wedge d \theta \wedge d
\phi,
\end{equation}
where $\tau = i(t- \frac{\pi}{2})$, the gauge is chosen such that
$A$ is regular at the south pole $(\tau = - \pi/2)$ of the
hemisphere $(0\geq \tau \geq - \pi/2)$. The gauge potential for
the north hemisphere will take the same form with a negative sign
in front of the constant term $\frac{2}{3}$. The sign change of
the potential is consistent with the time reversal, as we
mentioned earlier.

One can see that $A_{MNP}$ is subjected to a discontinuity at the
equator. Therefore, $A_{MNP}$ is not allowed to be the argument
for the instanton probability calculation in (11).

In order for the instanton approach to be valid, one has to use
the canonical conjugate representation. One can make a Fourier
transform of the wave function $\Psi (h_{ij}, A_{123})$ to get the
wave function $\Psi (h_{ij}, P^{123})$,
\begin{equation}
\Psi ((h_{ij}, P^{123}) = \frac{1}{2\pi} \int_{-\infty}^{\infty}
e^{iA_{123}P^{123}} \Psi(h_{ij}, A_{123}).
\end{equation}
where $P^{123}$ is the canonical momentum conjugate to $A_{123}$,
the only degree of freedom of the matter content under the
minisuperspace ansatz
\begin{equation}
P^{123} = \int_{\Sigma}\sqrt {- g_{11}} \left (- F^{0123} +
\frac{\sqrt{2}}{3(4!)} \eta^{0123m_5\cdots m_{11}}
F_{m_5m_6m_7m_8} A_{m_9m_{10} m_{11}}\right ) d^{10}x,
\end{equation}
where $\Sigma$ denotes the 10-dimensional surface $ t = const$.
The quantum transition should occur at the equator $\chi = \pi/2$.
However, the calculation at $\tau =0$ or $t = \pi/2$ is simpler.
Apparently, the result does not depend on the choice of the
equator (this has been confirmed), since all equators are
congruent for the round $S_4$ sphere. Strictly speaking, one
cannot use $A_{123}$ as the argument of the wave function without
gauge condition. The only meaning of this argument is its flux at
the surface. We shall not use the wave function $\Psi (h_{ij},
A_{123})$ anyway.

The discontinuity occurred at the equator instanton is thus
avoided using the momentum representation, although it is due to
the two distinct patches covering the whole sphere and can be
glued through a gauge transformation. At the $WKB$ level, the
Fourier transform of the wave function is equivalent to the
Legendre transform of the action. The Legendre transform has
introduced an extra contribution  $-2 A_{123}P^{123}$ to the
Euclidean action, where all quantities are in the Euclidean
version, and the factor 2 is due to the two sides of the equator
in the adjoining. Then the effective action becomes
\begin{equation}
I_{effect} = - \frac{2}{3} \kappa^2 V_7V_4.
\end{equation}

If we consider the quantum transition to occur at the equator of
$S_7$ instead, using the same argument, then it turns out that the
corresponding canonical momentum using the time coordinate in
$S_7$ vanishes, and the effective action should be the negative of
(13), taking account of the sign of the factor $\sqrt{g_{11}}$ in
the action (12).

Since the creation probability is the exponential to the negative
of the Euclidean action, the probability of creating a
7-dimensional macroscopic universe is exponentially suppressed
relative to that of the 4-dimensional case.

In the classical framework, the $S_7$ factor space in the
Freund-Rubin model can be replaced by $S_2\times S_5$, $S_2 \times
S_2 \times S_3$, $S_4 \times S_3$ or other Einstein spaces.
However, all these product spaces have volumes smaller than that
of $S_7$. It would lead to an exponential suppression of the
creation probability. Therefore, the internal space must be the
round $S_7$ space.

Now we consider the Englert model [5]. Then, in addition to the
components of the space $S_4$ in (7), the $F_{mnpq}$ components of
the $S_7$ space can be non-vanishing and satisfying
\begin{equation}
F^{mnpq}_{\;\;\;\;\;\;\;;m} = \left [
\frac{\sqrt{2}}{(4!)\sqrt{g_7}} \right ] \kappa \epsilon^{npqrstu}
F_{rstu}.
\end{equation}

Two nontrivial solutions are
\begin{equation}
F_{mnpq} = \frac{4}{\kappa} \partial_{[m} S^{\pm}_{npq]},
\end{equation}
where $S^{\pm}_{ mnp}= S^{\pm}_{[mnp]}$ are  the two torsion
tensors which can flatten the $S_7$ space in the Cartan-Schouten
sense [9]
\begin{equation}
R^m_{\;\;npq}\{\Gamma^r_{st} + S^r_{st}\} = 0,
\end{equation}
where $+\; (-)$ is for the case $\kappa > 0\; (\kappa < 0)$. It is
noted that $S_7$ is the only compact manifold to allow this, apart
from group manifolds. The potential can be chosen as
\begin{equation}
A_{mnp} = \frac{1}{6\kappa} S^\pm_{mnp}.
\end{equation}
The anisotropic cosmological constants are $\Lambda_7 =
3\kappa^2/4$ and $\Lambda_4=-5\kappa^2/4$.

The tensor $A_{mnp}$ satisfies the gauge condition
$A^{mnp}_{\;\;\;\; ;p} = 0$. The following properties of the
torsion tensor will be used in later calculations
\begin{equation}
S^{tr}_{\;\;\;\; m}S_{trn} =  \frac{3}{4} \kappa^2 g_{mn},
\end{equation}
\begin{equation}
S^{\pm \;mnp} = \mp \frac{2\sqrt{2}}{4!|\kappa | \sqrt{g_7}}
\epsilon^{mnpqrst}S^\pm_{ [rst,q]}.
\end{equation}

As in the Freund-Rubin model, before we take account of the
Legendre term, the Euclidean action of the Englert $AdS_4 \times
S_7$ space is
\begin{equation}
I =-\frac{1}{4} \kappa^2V_7V_4.
\end{equation}
After including the Legendre term the effective action becomes
\begin{equation}
I_{effect} =-\frac{2}{3} \kappa^2V_7V_4.
\end{equation}
It is surprising that after the long calculation, the effective
action remains the same as that in the Freund-Rubin case.

If the quantum transition occurred at an equator of the $S_7$
space, one has to include the Legendre terms correspondingly. In
contrast to the Freund-Rubin model, the canonical momenta do not
vanish. Fortunately, due to the symmetries of the torsion tensor,
the sum of the $C^3_6 = 20$  Legendre terms cancel exactly. The
action is the negative of that in (24)
\begin{equation}
I_{effect} =\frac{1}{4} \kappa^2V_7V_4.
\end{equation}
Again, comparing the results of (25) and (26), one can conclude
that the universe we are living is most likely 4-dimensional.

In the Freund-Rubin model, the $S_7$ factor space can be replaced
by a general Einstein space with the same cosmological constant
$\Lambda_7$. The Awada-Duff-Pope model [6] is most interesting.
The round 7-sphere is replaced by a squashed one, so that the
$N=8$ supersymmetry breaks down to $N = 1$.  As far as the
scenario of the quantum creation is concerned , the argument for
the Freund-Rubin model remains intact, the only alternations are
that the quantum transition should occur at one of its stationary
equators and $V_7$ should be the volume of the squashed 7-sphere.

There is no supersymmetry in the Englert model [5]. Englert,
Rooman and Spindel also discussed the model with a squashed $S_7$
factor space [7]. Here the  $A$ components in the $S_7$ space are
proportional to the torsion which renders the squashed sphere
Ricci-flat, instead. It is believed that our conclusion should
remain the same.

The right configuration for the wave function  has also been
chosen in the problem  of quantum creation of magnetic and
electric black holes [10].  If one considers the quantum creation
of a general charged and rotating black holes, this point is even
more critical. It is become so acute that unless the right
configuration is used, one even cannot find a seed constrained
instanton [11].

Many previous studies on dimensionality have essentially been
restricted to the classical framework. For $d=11$ supergravity,
there is no way to discriminate the $d=4$ and $d= 7$ macroscopic
universes in the classical framework, as in other similar but more
artificial models. This discrimination can be realized only
through quantum cosmology.

\vspace*{0.1in}

\bf References:

\vspace*{0.1in}
\rm

1. Z.C. Wu, \it Phys. Lett. \bf B\rm\underline{146}, 307 (1984).
X.M. Hu and Z.C. Wu, \it Phys. Lett. \bf B\rm\underline{149}, 87
(1984);  \it Phys. Lett. \bf B\rm\underline{155}, 237 (1985); \it
Phys. Lett. \bf B\rm\underline{182}, 305 (1986).

2. J.B. Hartle and S.W. Hawking, \it Phys. Rev. \rm \bf D\rm
\underline{28}, 2960 (1983).

3. E. Cremmer, B. Julia and J. Scherk, \it Phys. Lett. \bf
B\rm\underline{76}, 409 (1978).

4. G.O. Freund and M.A. Rubin, \it Phys. Lett. \bf
B\rm\underline{97}, 233 (1980).

5. F. Englert, \it Phys. Lett. \bf B\rm\underline{119}, 339
(1982).

6. M.A. Awada, M.J. Duff and C.N. Pope, \it Phys. Rev. Lett.
\rm\underline{50}, 294 (1983).

7. F. Englert, M. Rooman and P. Spindel, \it Phys. Lett. \bf
B\rm\underline{127}, 47 (1983).

8. J.J. Halliwell and S.W. Hawking, \it Phys. Rev. \rm \bf
D\rm\underline{31}, 346 (1985)

9. E. Cartan and J.A. Schouten, \it Proc. K. Acad. Wet. Amsterdam
\rm\underline{29}, 933 (1926).

10. S.W. Hawking and S.F. Ross, \it Phys. Rev. \rm \bf
D\rm\underline{52}, 5865 (1995). R.B. Mann and S.F. Ross, \it
Phys. Rev. \rm \bf D\rm\underline{52}, 2254 (1995).

11. Z.C. Wu, \it Int. J. Mod. Phys. \rm \bf D\rm \underline{6},
199 (1997), gr-qc/9801020; \it Phys. Lett. \bf B\rm
\underline{445}, 174 (1998), gr-qc/9810012. .

\end{document}